\documentclass[12pt]{article}
\usepackage[a4paper,margin=2cm]{geometry}
\usepackage{amsmath}
\usepackage{type1cm}
\usepackage{graphicx}
\usepackage{url} 
\usepackage{changepage}
\usepackage{textcomp,marvosym}
\usepackage{fixltx2e}
\usepackage{amsmath,amssymb}
\usepackage{nameref,hyperref}
\usepackage{authblk}
\usepackage{algorithm}
\usepackage{algpseudocode}
\usepackage{dsfont}
\usepackage{xcolor}
\usepackage{enumitem}
\usepackage{caption,setspace}
\usepackage{pdflscape}
\usepackage{graphicx}
\usepackage{placeins}
\usepackage{mathtools}
\usepackage{appendix}
\usepackage{extarrows}
\usepackage{natbib}
\usepackage{amsmath, amssymb, graphicx, psfrag}
\usepackage{color}
\usepackage{url}
\usepackage{lipsum}
\usepackage{physics}
\usepackage{siunitx}
\usepackage{bm}

\newcommand{\dydx}[2]{\frac{\text{d} #1}{\text{d} #2}}
\newcommand{\ddydx}[2]{\frac{\text{d}^2 #1}{\text{d} {#2}^2}}

\renewcommand{\eqref}[1]{Equation~(\ref{#1})}

\newcommand{\CondE}[2]{\mathbb{E}\left[#1 \mid #2\right]}

\newcommand{\CondPDF}[2]{\pi(#1 \mid #2)}
\newcommand{\approxCondPDF}[2]{\tilde{\pi}(#1 \mid #2)}

\newcommand{\like}[2]{\mathcal{L}(#2 ; #1)}
\newcommand{\genlike}[3]{\tilde{L}_{#3}(#2 ; #1)}

\newcommand{\loss}[2]{{\ell_{#1}(#2)}}

\newcommand{\V}[1]{\text{Var}\left[#1\right]}

\newcommand{\bvec}[1]{\mathbf{#1}}
\newcommand{\ind}[2]{\mathds{1}_{#1}\left(#2\right)}

\DeclareMathOperator*{\argmax}{argmax}
\DeclareMathOperator*{\argmin}{argmin}

\newcommand{\paramvec}{\boldsymbol{\theta}}

\newcommand{\param}{\theta}
\newcommand{\paramspace}{\boldsymbol{\Theta}}

\newcommand{\indicator}[2]{\mathds{1}_{#1}(#2)}

\begin{document}

\title{Generalised likelihood profiles for models with intractable likelihoods}

\author[1,2]{David~J.~Warne\footnote{To whom correspondence should be addressed. E-mail: david.warne@qut.edu.au}}
\author[3]{Oliver~J. Maclaren}
\author[1]{Elliot~J. Carr}
\author[1,2]{Matthew~J.~Simpson}
\author[1,2]{Christopher Drovandi}

\affil[1]{School of Mathematical Sciences, Queensland University of Technology, Brisbane, Australia}
\affil[2]{Centre for Data Science, Queensland University of Technology, Brisbane, Australia}
\affil[3]{Department of Engineering Science, University of Auckland, Auckland, New Zealand}
\date{\today}
\maketitle

\begin{abstract}
Likelihood profiling is an efficient and powerful frequentist approach for parameter estimation, uncertainty quantification and practical identifiablity analysis. Unfortunately, these methods cannot be easily applied for stochastic models without a tractable likelihood function. Such models are typical in many fields of science, rendering these classical approaches impractical in these settings. To address this limitation, we develop a new approach to generalising the methods of likelihood profiling for situations when the likelihood cannot be evaluated but stochastic simulations of the assumed data generating process are possible. Our approach is based upon recasting developments from generalised Bayesian inference into a frequentist setting. We derive a method for constructing generalised likelihood profiles and calibrating these profiles to achieve desired frequentist coverage for a given coverage level. We demonstrate the performance of our method on realistic examples from the literature and highlight the capability of our approach for the purpose of practical identifability analysis for models with intractable likelihoods.    
\end{abstract}

\section{Introduction}

Uncertainty quantification is an essential tool for parameter estimation and prediction in all fields sciences~\citep{Box1976,Psaros2023,Volodina2021}. In most settings, statistical analysis relies on a sensible likelihood function to be constructed for the model of interest. When this likelihood function is tractable, a variety of standard tools are available to perform statistical analysis, the most appropriate of which will be problem dependent. In the classical setting, maximum likelihood estimation is typically used for parameter estimates and uncertainty is quantified through an estimate of the confidence sets~\citep{Casella2002,pawitan2001all,Wasserman2014}. For Bayesian analysis, both parameter estimation and uncertainty quantification are performed together via sampling of the posterior distribution~\citep{Gelman2014,McElreath2020}.



 However, stochastic models are routinely relied upon to describe complex phenomena in many scientific disciplines including,  astronomy~\citep{Lemos2023,Zhao2022}, systems biology~\citep{Browning2020,Szekely2014,Wilkinson2018}, climatology \citep{Palmer2019,Peleg2019}, ecology~\citep{Adams2020,Black2012,Shoemaker2019}, epidemiology~\citep{Chinazzi2020,Li2020}, geology \citep{Wang2022,Vasylkivska2021}, and particle physics~\citep{Brehmer2021,Kiss2023}. Since such models rarely have tractable likelihood functions, methods that avoid direct evaluation of the likelihood function are required. In Bayesian settings, approximate Bayesian computation~\citep{Sisson2018,Beaumont2010}, Bayesian synthetic likelihood~\citep{Price2017}, pseudo-marginal methods~\citep{Warne2020,Andrieu2010}, and more recently deep learning methods~\citep{Cranmer2020,Papamakarios2019,Kelly2023} are frequently utilised to perform sampling of the posterior or an approximation thereof. There are also classical methods that use alternative estimators to the maximum likelihood estimator or methods that are simulation-based, such as $M$-estimators~\citep{Huber2009,Shapiro2000}, and bootstrap sampling~\citep{Efron1979,Kleiner2014}. However, many classical approaches rely on large data sample asymptotics to obtain confidence sets. Such requirements are almost never satisfied in practice, and therefore the resulting estimates are often overconfident~\citep{sprott2008statistical,Pawitan2000}. Furthermore, these standard approaches completely fail when parameter identifiability cannot be guaranteed~\citep{Raue2009,simpson2020practical}.
 
 Likelihood profiles provide a means of estimating parameters and confidence sets that i) typically have better small sample behaviour than Wald-style confidence intervals; ii) can handle asymmetric confidence sets; and iii) can be used in cases when there is practical parameter non-identifiability~\citep{Browning2020,Hines2014,Murphy2022,Raue2009,Roosa2019}. In this work, we extend the likelihood profiling approach into the simulation-based inference setting where the likelihood cannot be evaluated explicitly. We construct a new estimator, called the maximum generalised likelihood estimator (MGLE). This estimator is based upon a generalisation of the likelihood in the sense of generalised Bayesian inference~\citep{Bissiri2016,Matsubara2022}. The approach involves construction of generalised likelihood profiles and applies bootstrap samples to calibrate the profiles to an appropriate scale and obtain confidence sets through setting thresholds~\citep{Pawitan2000,pawitan2001all}. Demonstrations of our approach are provided for several intractable stochastic models that arise in ecological and biological applications. Compared with standard bootstrap sampling, the generalised likelihood profile has improved coverage characteristics and may be employed  to assess practical identifiability for parameters of interest.              



\section{Background}
\label{sec:bg}

A common approach to frequentist statistical inference for many problems is to use the likelihood function to form both point and interval estimates \citep{cox2006principles,pawitan2001all,sprott2008statistical}. As is well known, the likelihood function treats an assumed family of generative probability models for the data as a function of the parameter for fixed data. The profile likelihood is defined in terms of the likelihood function and allows for targeting parameters of interest \citep{cox2006principles,pawitan2001all,pace1997principles,sprott2008statistical}. However, while there have been attempts to develop axiomatic pure likelihood approaches to inference \citep{edwards1984likelihood,Royall2017statistical}, the primary frequentist justification of likelihood-based inference, besides its ease of use and parameterisation invariance, is its good repeated sampling performance for many problems. This performance includes asymptotic optimality arguments that hold under certain restricted conditions \citep{lecam1990maximum, severini2000likelihood}. Parameterisation invariance and natural respect for parameter constraints motivate using likelihood and profile likelihood for inference in finite samples \citep{pawitan2001all,pace1997principles,sprott2008statistical}. Importantly, in addition to its good properties for standard problems, profile likelihood is also a reliable tool for structural and practical identifiability analysis \citep{Raue2009, simpson2020practical}.

Despite good reasons for using a likelihood-based approach to frequentist inference, there is no necessity within frequentist inference to base estimation on the model likelihood function. This recognition and a desire to produce robust or more convenient estimation procedures has led to the development of numerous approaches to frequentist inference that do not use the likelihood. Many of these are natural generalisations of likelihood-based inference, particularly $M$-estimation (for `Maximum-likelihood-like estimation') and $Z$-estimation introduced by \citet{Huber1964} and the closely related theories of estimating equations~\citep{Durbin1960,Godambe1960}, inference functions~\citep{McLeish1988,Lindsay2003}, quasi-likelihood \citep{McCullagh1989,Wedderburn1974}, and minimum distance estimation~\citep{Parr1980,Wolfowitz1957}. 

Many of the above methods generalise the theory of likelihood-based point estimation and the associated asymptotic theory of `Wald-style' \citep{Wald1939,Shapiro2000} intervals based on the sampling distribution of a point estimator. The bootstrap~\citep{Efron1979} is a natural, more sophisticated form of constructing intervals from point estimates and their sampling distributions. An approach like the bootstrap also fits naturally with simulation-based inference: given an $M$-estimator or similar, we can, for example, simulate the estimator's sampling distribution by using a plug-in best estimate of the parameters. We can then use this distribution to form bootstrap confidence intervals. Related ideas on simulation-based inference appear in the econometrics literature~\citep{Contoyannis2004,Geweke2000,Winker2007,gourieroux1996simulation}. However, confidence intervals in such approaches are still typically based on a point estimator and its sampling distribution. This means neglecting some of the comparatively desirable properties of full-likelihood-based methods in small samples \citep{Meeker1995,Pawitan2000}. Methods analogous to those that use more information from the full likelihood than just the maximum likelihood estimate, including profile likelihood inference, have received less attention. Notable exceptions include the concept of empirical likelihood \citep{Owen1988,Owen1990},  profile estimating equations~\citep{Bellio2008,Li1994,Liang1995}, and recent work by~\citet{Dalmasso2020} on approximate computation via odds ratio estimation (ACORE). We also recently considered profiling a special (approximately Gaussian) class of `surrogate' likelihood functions that can be applied to simulation-based inference when an approximation to the simulation model is available \citep{simpson2022reliable}. In addition, the work of \cite{Ionides2017Monte} on simulation-based profile likelihood inference relates closely to ours. However, they look at calibrating Monte Carlo estimates of the implicit likelihood function of the model. In contrast, we consider the situation of inference using a deterministic loss derived from properties of the model. In this sense our approach results in a different but likelihood-like estimation function.

Given this context, here we present a simple approach to simulation-based frequentist inference in the style of $M$-estimation that more closely follows the methodology of profile likelihood. In particular, we consider the so-called generalised likelihood function as it arises in generalised Bayesian inference \citep{Bissiri2016,Matsubara2022}. Though in the broader context outlined above, such a function is just one way of generalising a likelihood function, it has some appealing properties motivating its use in the Bayesian literature \citep{matsubara2022generalised}. Here we explore its use in the frequentist context and evaluate its performance in terms of its ability to achieve good interval coverage and precision. We propose a simple bootstrap procedure to first calibrate the generalised likelihood function's single unknown parameter. We then use the resulting function like a standard likelihood function to carry out likelihood-based-like inference, aiming to take advantage of its potential for improvements on Wald-style intervals. We focus, in particular, on profiling generalised likelihood functions to obtain confidence intervals for interest parameters.

\section{Methods}
In this section, we present our approach for parameter and confidence set estimation when the likelihood is intractable. We let $\paramvec = (\param_1,\param_2,\ldots,\param_d)^\text{T} \in \paramspace \subseteq \mathbb{R}^d$ be a vector of $d \geq 1$ model parameters and let $\bvec{y} = (y_1,y_2,\ldots, y_n)^\text{T}$ be the data consisting of $n \geq 1$ points. For the setting of interest, the likelihood function for the model $\like{\bvec{y}}{\paramvec} = \CondPDF{\bvec{y}}{\paramvec}$ cannot be evaluated directly. However, we assume: 1) it is possible to generate exact realisations from the model, $\bvec{y}_s \sim \CondPDF{\cdot}{\paramvec}$, and 2) that we can evaluate an appropriate loss function $\ell_{\bvec{y}} : \paramspace \rightarrow \mathbb{R}^+$ directly. We will discuss different options for this loss function throughout this section, however, the main qualitative conditions we require are 1) as $n \to \infty$ the minimum value of the loss function is in the neighbourhood of the true parameter $\paramvec_0$  that generated $\bvec{y}$, and 2) the loss function increases away from this neighbourhood of $\paramvec_0$.

\subsection{The maximum generalised likelihood esitmator}

For a suitable loss function $\loss{\bvec{y}}{\paramvec}$, and tuning parameter $\delta > 0$, we define the function
\begin{equation}
\genlike{\bvec{y}}{\paramvec}{\delta} = e^{-\delta \loss{\bvec{y}}{\paramvec}}.
\label{eq:genlike}
\end{equation}
Note that if the loss function is chosen to be $\loss{\bvec{y}}{\paramvec} = -\delta^{-1}\log \like{\bvec{y}}{\paramvec}$, then we arrive at $\genlike{\bvec{y}}{\paramvec}{\delta} = \like{\bvec{y}}{\paramvec}$ as a special case. In this sense, we refer to \eqref{eq:genlike} as the generalised likelihood function in the same way that the generalised posterior distribution is defined by ~\citet{Bissiri2016} and \citet{Matsubara2022}. 

Given this generalised likelihood function, we can consider parameter estimation by using a maximum generalised likelihood estimator (MGLE),
\begin{align}
\hat{\paramvec}_{\text{MGLE}} = \argmax_{\paramvec \in \paramspace} \genlike{\bvec{y}}{\paramvec}{\delta}= \argmax_{\paramvec \in \paramspace} e^{-\delta \loss{\bvec{y}}{\paramvec}} = \argmin_{\paramvec \in \paramspace} \delta \loss{\bvec{y}}{\paramvec}.
\label{eq:mgle}
\end{align}
Importantly, we note that the distribution of the estimator, $\hat{\paramvec}_{\text{MGLE}}$, for random data is unaffected by the value of tuning parameter $\delta$. Hence $\delta$ is a arbitrary positive number for the purpose of point parameter estimation. However, as we demonstrate in section~\ref{sec:delta_choice}, the inclusion of $\delta$ enables us to rescale $\genlike{\bvec{y}}{\paramvec}{\delta}$ appropriately to obtain confidence intervals at a prescribed targeted coverage in a similar way to the application of thresholds to likelihood profiles~\citep{Raue2009,pawitan2001all}.

Assuming a fixed value for $\delta$, then for any injective function of the parameters, $\boldsymbol{\tau} = f(\paramvec)$ with $\boldsymbol{\tau} \in \mathbb{R}^q$ for $q \geq 1$, we share the invariance property of the maximum likelihood, that is $\hat{\boldsymbol{\tau}}_{\text{MGLE}} = f(\hat{\paramvec}_{\text{MGLE}})$. This can be shown by following arguments identical to that of \citet{Zehna1966} in the context of the standard maximum likelihood estimator. We can construct a generalised likelihood induced by the transformation $\boldsymbol{\tau} = f(\paramvec)$,
\begin{equation*}
\genlike{\bvec{y}}{\boldsymbol{\tau}}{\delta} = \sup_{\paramvec \in \paramspace : f(\paramvec) = \boldsymbol{\tau}} \genlike{\bvec{y}}{\paramvec}{\delta}.
\end{equation*} 
By definition, we have that $\genlike{\bvec{y}}{\boldsymbol{\tau}}{\delta} \leq \sup_{\paramvec \in \paramspace} \genlike{\bvec{y}}{\paramvec}{\delta} = \genlike{\bvec{y}}{\hat{\paramvec}_{\text{MGLE}}}{\delta}$, then it follows that  $\sup_{\boldsymbol{\tau} \in \mathbb{R}^q} \genlike{\bvec{y}}{\boldsymbol{\tau}}{\delta} = \genlike{\bvec{y}}{\hat{\boldsymbol{\tau}}_{\text{MGLE}}}{\delta} = \genlike{\bvec{y}}{\hat{\paramvec}_{\text{MGLE}}}{\delta}$. Furthermore, our definition in fact ensures the generalised likelihood is fully invariant under bijective transformations as the above reduces to $\genlike{\bvec{y}}{\boldsymbol{\tau}}{\delta}=\genlike{\bvec{y}}{\paramvec}{\delta}$ when $\paramvec$ is a unique inverse of $\boldsymbol{\tau}$.

 While there is a wide variety of loss functions for which $\genlike{\bvec{y}}{\paramvec}{\delta}$ is likelihood-like in the sense described in Section~\ref{sec:bg}, we restrict ourselves to loss functions that fall within the family of minimum kernel discrepancy estimators~\citep{Oates2022}. By doing this we ensure asymptotic consistency of the MGLE. 

\subsection{Generalised likelihood profiles and confidence sets}
 \label{sec:glp}
If we partition the parameter vector, $\paramvec$ into a vector of interest parameters, $\boldsymbol{\phi} \in \boldsymbol{\Phi}$ and nuisance parameters $\boldsymbol{\psi} \in \boldsymbol{\Psi}$ such that $\paramvec = (\boldsymbol{\phi},\boldsymbol{\psi})$ and $\paramspace = \boldsymbol{\Phi} \times \boldsymbol{\Psi}$, then we can define the generalised likelihood profile of the interest parameters by optimising out the nuisance parameters using
\begin{equation}
\genlike{\bvec{y}}{\boldsymbol{\phi}}{\delta} =  \sup_{\boldsymbol{\psi} \in \boldsymbol{\Psi}} \genlike{\bvec{y}}{(\boldsymbol{\phi},\boldsymbol{\psi})}{\delta}.
\label{eq:genlikeprof}
\end{equation}
Typically, the profile will be evaluated for a grid of $M$ points $\{\boldsymbol{\phi}^j\}_{j=1}^M$. Just as in standard likelihood profiles, the accuracy of the profile relies on the resolution of the grid which may be difficult to assess \textit{a priori}~\citep{Simpson2021}. Thus, adaptive mesh refinements may be necessary in general. For the numerical examples in this work, we focus on $1$-dimensional profiles only and use regular grids in each axis. For example, if the $i$th dimension of $\paramvec$ is of interest, then we set $\boldsymbol{\phi} = \param_i$ and $\boldsymbol{\psi} = (\param_1,\ldots,\param_{i-1},\param_{i+1},\ldots,\param_d)$, and define the grid over the domain $[L_i,U_i]$ as $\param_i^j = L_i + (j-1)\Delta\param_i$ for $j \in [1,2,\ldots,M]$ with $\Delta\param_i = (U_i - L_i)/(M-1)$. Our methods, however, are completely applicable to higher-dimensional parameters of interest.


While the MGLE does not depend on $\delta$, construction of a meaningful generalised likelihood profile will require \eqref{eq:genlikeprof} to be evaluated, thus the tuning parameter $\delta$ is required. This leads to the question of how this parameter should be chosen. We propose that $\delta$ be calibrated in such a way that the confidence set obtained by the generalised likelihood profile has approximately the correct coverage in a frequentist setting given the usual likelihood ratio-based threshold from Wilks' theorem~\citep{pawitan2001all}.

We proceed as follows, given $\hat{\paramvec}_{\text{MGLE}}$ as given in \eqref{eq:mgle}, generate $K$ simulated bootstrap datasets from the model,
\begin{equation*}
\bvec{y}_k \sim \CondPDF{\cdot}{\hat{\paramvec}_{\text{MGLE}}}, \quad k = 1,2,\ldots, K.
\end{equation*} 
Then, for a given value of $\delta$, we can estimate the coverage at $\hat{\paramvec}_{\text{MGLE}}$. We do this by estimating the MGLE and generalised likelihood profiles, that is, we obtain, for $k \in [1,2,\ldots K]$,
\begin{equation*}
		\hat{\paramvec}_{\text{MGLE}}^k = \argmin_{\paramvec \in \paramspace} \loss{\bvec{y}_k}{\paramvec} \quad\text{and} \quad\{\genlike{\bvec{y}_k}{\boldsymbol{\phi}_k^j}{\delta}\}_{j=1}^{M}, \\
\end{equation*}
where $\{\boldsymbol{\phi}_k^j\}_{j=1}^M$ is the grid for the parameters of interest and $\genlike{\bvec{y}_k}{\boldsymbol{\phi}_k^j}{\delta} = \sup_{\boldsymbol{\psi} \in \boldsymbol{\Psi}}\genlike{\bvec{y}_k}{(\boldsymbol{\phi}_k^j,\boldsymbol{\psi})}{\delta}$. The $(1-\alpha)100\%$ confidence set can then be determined by  $\boldsymbol{\Phi}_{(1-\alpha)100\% \text{CI}}^k = \{\boldsymbol{\phi}_k^j : \log \genlike{\bvec{y}_k}{\boldsymbol{\phi}_k^j}{\delta} > c_k \}$ with $c_k = -\delta \loss{\bvec{y}_k}{\hat{\paramvec}^k_{\text{MGLE}}} - \tau_\alpha$ and $\tau_\alpha$ is the appropriate cut-off at the $\alpha$ significance level. The empirical coverage can then be estimated by, $C_{\text{emp}}(\delta) = \frac{1}{K} \sum_{k=1}^K \indicator{\boldsymbol{\Phi}_{(1-\alpha)100\% \text{CI}}^k}{\hat{\boldsymbol{\phi}}_{\text{MGLE}}}$ where $\hat{\boldsymbol{\phi}}_{\text{MGLE}}$ are the parameters of interest at the MGLE.

Unfortunately, in practice we do not know the asymptotic distribution of the generalised likelihood equivalent of the likelihood ratio-test for a given $\delta$, and we cannot expect the usual threshold, $\tau_\alpha$, obtained through Wilks' theorem to yield the desired coverage properties. Therefore we have no analytical method to select $\tau_\alpha$. Instead, we optimise 
\begin{equation*}
	\delta^* = \argmin_{\delta \in \mathbb{R}^+} | C_{\text{emp}}(\delta) - (1-\alpha)|,
\end{equation*}
given the usual $\tau_\alpha = \frac{1}{2}F^{-1}_{\chi_{\dim(\phi)}^2}(1-\alpha)$ as per Wilks' theorem where $F^{-1}_{\chi_{\dim(\boldsymbol{\phi})}^2}(\cdot)$ denotes the quantile function of the $\chi^2$ distribution with degrees of freedom equal to the dimensionality of the interest parameters $\boldsymbol{\phi}$. Thus we arrive at the interpretation for $\delta$ as a tuning parameter to calibrate the generalised likelihood to the threshold given by Wilks' theorem.

\subsection{Calibration of $\delta$ tuning parameter}
Since the MGLE is invariant to the choice of $\delta$ (\eqref{eq:genlike}), then we have the property,
\begin{equation*}
\genlike{\bvec{y}}{\boldsymbol{\phi}}{\delta} = \sup_{\boldsymbol{\psi} \in \boldsymbol{\Psi}} e^{-\delta \loss{\bvec{y}}{(\boldsymbol{\phi},\boldsymbol{\psi})}} = \sup_{\boldsymbol{\psi} \in \boldsymbol{\Psi}} \left[e^{- \loss{\bvec{y}}{(\boldsymbol{\phi},\boldsymbol{\psi})}}\right]^\delta = \left[\sup_{\boldsymbol{\psi} \in \boldsymbol{\Psi}} e^{- \loss{\bvec{y}}{(\boldsymbol{\phi},\boldsymbol{\psi})}}\right]^\delta =  \left[\genlike{\bvec{y}}{\boldsymbol{\phi}}{1}\right]^\delta.
\end{equation*}
As a result, we need not recompute the generalised profiles as we optimise $\delta$ for any given target coverage at the MGLE. Instead we need only compute the profile of interest  once for each of the simulated datasets $\mathcal{Y} = [\bvec{y}_1,\bvec{y}_2, \ldots, \bvec{y}_K]$ with $\delta = 1$, then perform optimisation for $\delta$. To implement this, we define the function,
\begin{equation}
	C(\delta ;\tau_\alpha,\hat{\boldsymbol{\phi}}_{\text{MGLE}},\{\hat{\paramvec}^k_{\text{MGLE}}\}_{k=1}^K,\{\genlike{\bvec{y}_k}{\cdot}{1}\}_{k=1}^K) = \frac{1}{K}\sum_{k=1}^K \ind{[c_k,\infty)}{ \delta\log\genlike{\bvec{y}_k}{\hat{\boldsymbol{\phi}}_{\text{MGLE}}}{1}},
	\label{eq:deltacov}
\end{equation}  
with $c_k = -\tau_\alpha  -\delta\loss{\bvec{y}_k}{\hat{\paramvec}^k_{\text{MGLE}}}$ for each $k \in [1,2,\ldots, K]$.  We then define a step size $\Delta \delta$ and grid resolution $M_\delta$ and then find $\delta^* \in [0,\Delta \delta,2\Delta \delta,\ldots,M_\delta \Delta \delta]$ such that the absolute difference between \eqref{eq:deltacov} and the target coverage $1-\alpha$ is minimised. Of course, more sophisticated optimisation approaches could be used in this case, however, the computational cost of this grid search is negligible compared to the computational cost of optimisations required for the $K + 1$ MGLEs and generalised likelihood profiles for which we apply the interior-point method as implemented in MATLAB$^{\textregistered}$'s \texttt{fmincon}~\citep{Byrd2000}. This calibration process proceeds as given in Algorithm~\ref{alg:calibdelta}.

\begin{algorithm}
	\caption{Calibration process for generalised likelihood profiles.}
	\begin{algorithmic}
		\State Given data $\bvec{y}$, data generating process $\CondPDF{\bvec{y}}{\paramvec}$, loss function $\loss{\bvec{y}}{\paramvec}$, interest parameter $\boldsymbol{\phi}$, profile grid resolution $M$, and target significance level $\alpha \in (0, 1)$;
		\State Estimate MGLE, $\hat{\paramvec}_{\text{MGLE}} \leftarrow \argmin_{\paramvec \in \paramspace} \loss{\bvec{y}}{\paramvec}$;
		\State Estimate the profile $\genlike{\bvec{y}}{\cdot}{1}$ using \eqref{eq:genlikeprof} over $\boldsymbol{\Phi}$ using a grid of resolution $M$;
		\For{$k \in [1,2,\ldots,K]$}
		\State Generate simulated data $\bvec{y}_k \sim \CondPDF{\cdot}{\hat{\paramvec}_{\text{MGLE}}}$;
		\State Estimate MGLE, $\hat{\paramvec}^k_{\text{MGLE}} \leftarrow \argmin_{\paramvec \in \paramspace} \loss{\bvec{y}_k}{\paramvec}$;
		\State Estimate the profile $\genlike{\bvec{y}_k}{\cdot}{1}$ using \eqref{eq:genlikeprof} over $\boldsymbol{\Phi}$ using a grid of resolution $M$;
		\EndFor 
		\State Obtain threshold as per Wilks' theorem, $\tau_\alpha \leftarrow \frac{1}{2}F^{-1}_{\chi_{\dim(\phi)}^2}(1-\alpha)$;
		\State $\delta^* \leftarrow 0$, and $C^* \leftarrow 1$;
		\For{$\delta \in [\Delta \delta,2\Delta \delta,\ldots,M_\delta \Delta \delta]$}
		\State $C_\delta \leftarrow C(\delta ;\tau_\alpha,\hat{\boldsymbol{\phi}}_{\text{MGLE}},\{\hat{\paramvec}^k_{\text{MGLE}}\}_{k=1}^K,\{\genlike{\bvec{y}_k}{\cdot}{1}\}_{k=1}^K)$;
		\If{$|C_\delta - (1-\alpha)| < |C^* - (1-\alpha)| $}
		\State $C^* \leftarrow C_\delta$, and $\delta^* \leftarrow \delta$;
		\EndIf
		\EndFor
	\end{algorithmic}
\label{alg:calibdelta}
\end{algorithm}

Once Algorithm~\ref{alg:calibdelta} has been executed, we can obtain an approximation for the $(1-\alpha)100\%$ confidence set for parameters of interest under the real data $\bvec{y}$, that is $\boldsymbol{\Phi}_{(1-\alpha)100\% \text{CI}}^* = \{\boldsymbol{\phi}^j : \log \genlike{\bvec{y}}{\boldsymbol{\phi}^j}{\delta^*} > c^* \}$ with $c^* = -\delta^* \loss{\bvec{y}}{\hat{\paramvec}_{\text{MGLE}}} - \tau_\alpha$.

For the examples presented in this manuscript, we verify the empirical coverage properties hold independently of the data used as a part of the calibration process. To validate this, we consider $B$ simulated datasets under a set of known true parameter vector $\paramvec_{\text{true}}$, that is, $\bvec{y}_1, \bvec{y}_2, \ldots, \bvec{y}_B \sim \CondPDF{\cdot}{\paramvec_{\text{true}}}$. For each simulated dataset, $\bvec{y}_b$, we obtain an MGLE, $\hat{\paramvec}^b_{\text{MGLE}}$, and a calibrated profile, $\genlike{\bvec{y}_b}{\cdot}{\delta^*}$ where $\delta^*$ is the tuning parameter calibrated for $\bvec{y}$. The emprical coverage for $\boldsymbol{\phi}_{\text{true}}$ at significance level $\alpha$, $C_\alpha$, is then estimated using, 
\begin{equation}
	C_\alpha = \frac{1}{B}\sum_{b=1}^B \ind{[c_b^*,\infty)}{ \log\genlike{\bvec{y}_b}{\boldsymbol{\phi}_{\text{true}}}{\delta^*}},	
	\label{eq:cov_propr}
\end{equation}
where $c_b^* = -\tau_\alpha  -\delta^*\loss{\bvec{y}_b}{\hat{\paramvec}^b_{\text{MGLE}}}$ for each $b \in [1,2,\ldots, B]$.  Crucially, note that \eqref{eq:cov_propr} estimates the proper coverage for $\boldsymbol{\phi}_{\text{true}}$ rather than the bootstrap coverage for $\hat{\boldsymbol{\phi}}^b_{\text{MGLE}}$ that forms the basis of the calibration process. For our numerical experiments in Section~\ref{sec:results}, we estimate the proper coverage in this manner for several sets of $\paramvec_{\text{true}}$, but we do not do this exhaustively over all possible $\paramvec_{\text{true}}$ (See Discussion in Section~\ref{sec:discuss} for some possible ways to assess the coverage over all $\paramvec_{\text{true}}$). 

\section{Results}
\label{sec:results}

In this section, we demonstrate our generalised likelihood profile approach for parameter estimation, uncertainty quantification and assessment of practical identifiability. We first demonstrate the accuracy of the method using an intractable statistical model commonly used in ecology that has accurate approximations available. We then move to a stochastic model used in the study of stochastic advection-diffusion-reaction processes in biology based on the work of \citet{Carr2019}, \citet{Ellery2012} and \citet{Simpson2021}. The second represents a realistic use case for our approach since one can only obtain an analytical approximation to the moments of a particle lifetime.  

%
%
\subsection{Conway--Maxwell--Poisson model}

The Conway-Maxwell-Poisson (CMP) model \citep{conway1962queuing} is an extension of the Poisson distribution that can accommodate underdispersed and overdispersed count data. Applications for this model are especially useful in ecological studies~\citep{Kolb2020,Lynch2014,Wu2013}. The probability mass function for a single observation $y$ given parameter $\paramvec = (\lambda, \nu)^\text{T}$ is:
\begin{equation*}
    \CondPDF{y}{\paramvec} = \frac{\approxCondPDF{y}{\paramvec}}{ Z_{\paramvec}}, \text{where }\approxCondPDF{y}{\paramvec} = \frac{\lambda^y}{(y!)^{\nu}},
\end{equation*}
with $\lambda > 0$ and $\nu > 0$. The normalising constant $Z_{\paramvec} = \sum_{y=0}^\infty \approxCondPDF{y}{\paramvec}$ does not have a closed form expression, except for certain special cases.  However, the normalising constant can be computed to high accuracy with little computational effort~\citep{Sellers2010}, and facilitates comparison of our proposed generalised likelihood method with an accurate approximation of the true profile likelihood.

Following \citet{matsubara2022generalised}, we avoid evaluation of the intractable normalising constant $Z_{\boldsymbol{\theta}}$ using the discrete Fisher divergence (DFD) as the loss function~\citep{Xu2022}.  The discrete Fisher divergence,  in the one-dimensional data setting as considered here, between the statistical model conditioned on $\paramvec$, $\CondPDF{\cdot}{\paramvec}$, and the empirical distribution of the data, $\pi_n$, is given by
\begin{align}
   \loss{\bvec{y}}{\paramvec} = \mbox{DFD}(\CondPDF{\cdot}{\paramvec}|| \pi_n) &=  \frac{1}{n} \sum_{i=1}^n \left( \frac{\CondPDF{y_i^-}{\paramvec}}{\CondPDF{y_i}{\paramvec}}  \right)^2 - 2 \left( \frac{\CondPDF{y_i}{\paramvec}}{\CondPDF{y_i^+}{\paramvec}}  \right). 
   \label{eq:dfd}
\end{align}
Note that the intractable normalising constants cancel in the ratio. For this example, we use the notation $y_i^+ = y_i + 1$ and $y_i^- = y_i-1$, unless $y_i=0$ in which case we set $y_i^- = \mathrm{max} \{y_k\}_{k=1}^n$, i.e.\ the maximum value of the dataset.

\citet{matsubara2022generalised} embed the DFD within a generalised Bayes framework to conduct approximate Bayesian inferences without invoking the $\paramvec$-dependent normalisation constant.  In order to calibrate the scaling parameter, \citet{matsubara2022generalised} firstly generate $B$ bootstrap estimates of $\paramvec$ from datasets that are generated from the model with the value of $\paramvec$ that minimises the DFD based on the observed dataset. Then, \citet{matsubara2022generalised} compute the scaling parameter that minimises the Fisher divergence between the generalised posterior and the empirical bootstrap sample.  Here we show that $\delta$ can instead be calibrated using our profile likelihood approach.

We consider $n = 2000$ samples from the CMP model, $y_1, y_2, \ldots, y_n \sim \CondPDF{\cdot}{\lambda,\nu}$ ,  with $\lambda = 4$ and $\nu = 2$. The MGLE is then obtained through numerical optimization of \eqref{eq:mgle} with the DFD (\eqref{eq:dfd}) substituted for the loss function. Univariate generalised likelihood profiles are generated over uniform discretisations of $M = 100$ points over sensible ranges for the two interest parameters,   $1 \leq \lambda \leq 20$, and $1  \leq \nu \leq 8$, respectively. The procedure in from Section \ref{sec:glp} is applied  using $K = 100$ bootstrap samples, then calibrate profiles with significance level $\alpha = 0.05$ and threshold of $\tau_\alpha = -1.92$ to obtain $95\%$ confidence intervals. In addition, standard MLE and likelihood profiles are estimated for comparison using the accurate numerical approximation for the normalising constant $Z_{\boldsymbol{\theta}}$.

\begin{figure}[h]
	\centering
	\includegraphics[width=0.85\linewidth]{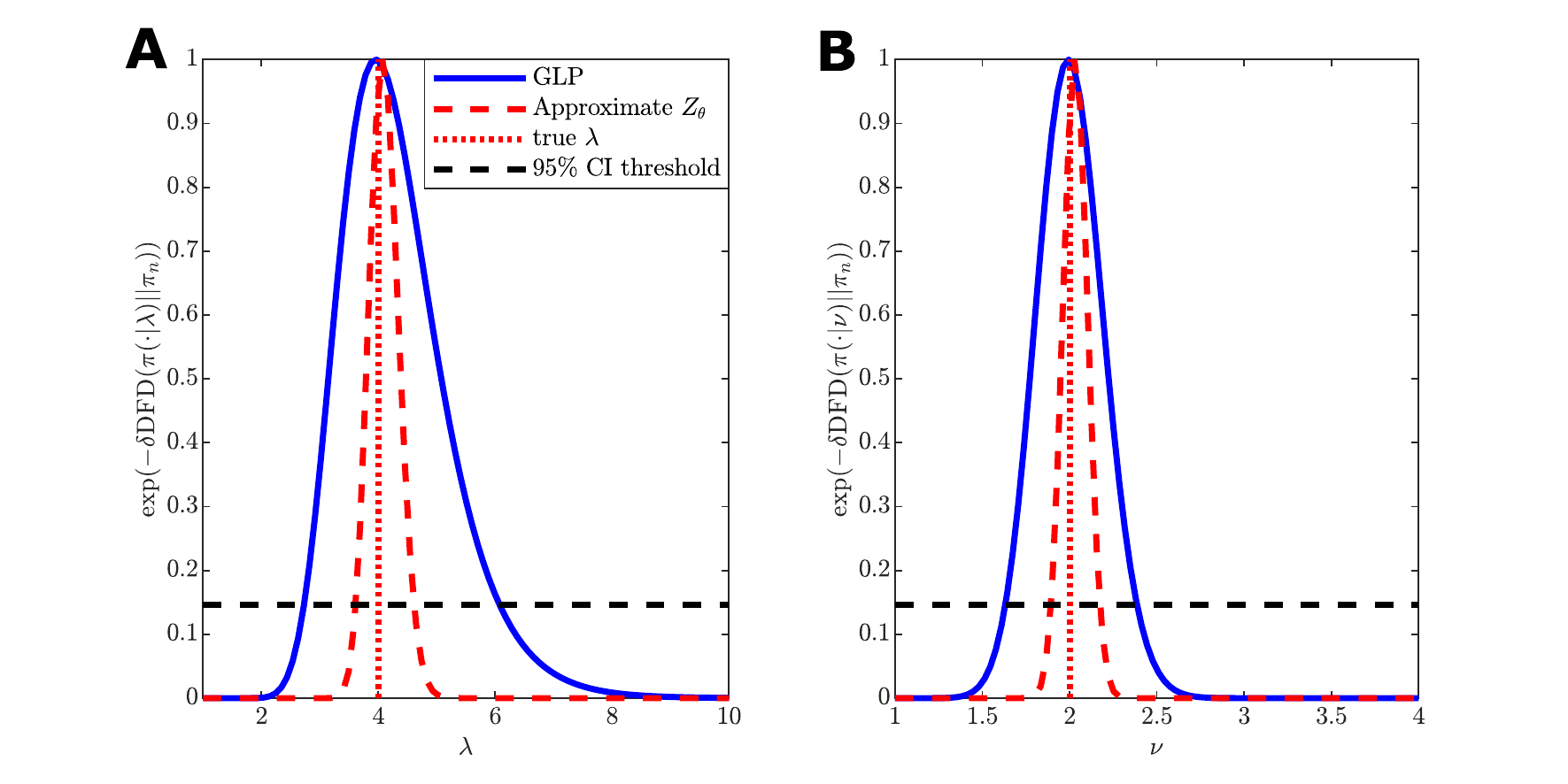}
	\caption{Comparison of a generalised likelihood profile (solid blue) with a standard likelihood profile (dashed red) using an accurate numerical approximation for $Z_{\boldsymbol{\theta}}$ for the CMP model. The 95\% confidence interval threshold of $\tau_{0.05} = -1.92$ is indicated (dashed black). The true parameters used to generate the data are $\lambda = 4$, and $\nu = 2$ (dotted red).}
	\label{fig:prcmp}
\end{figure}

\begin{figure}[h]
	\centering
	\includegraphics[width=0.85\linewidth]{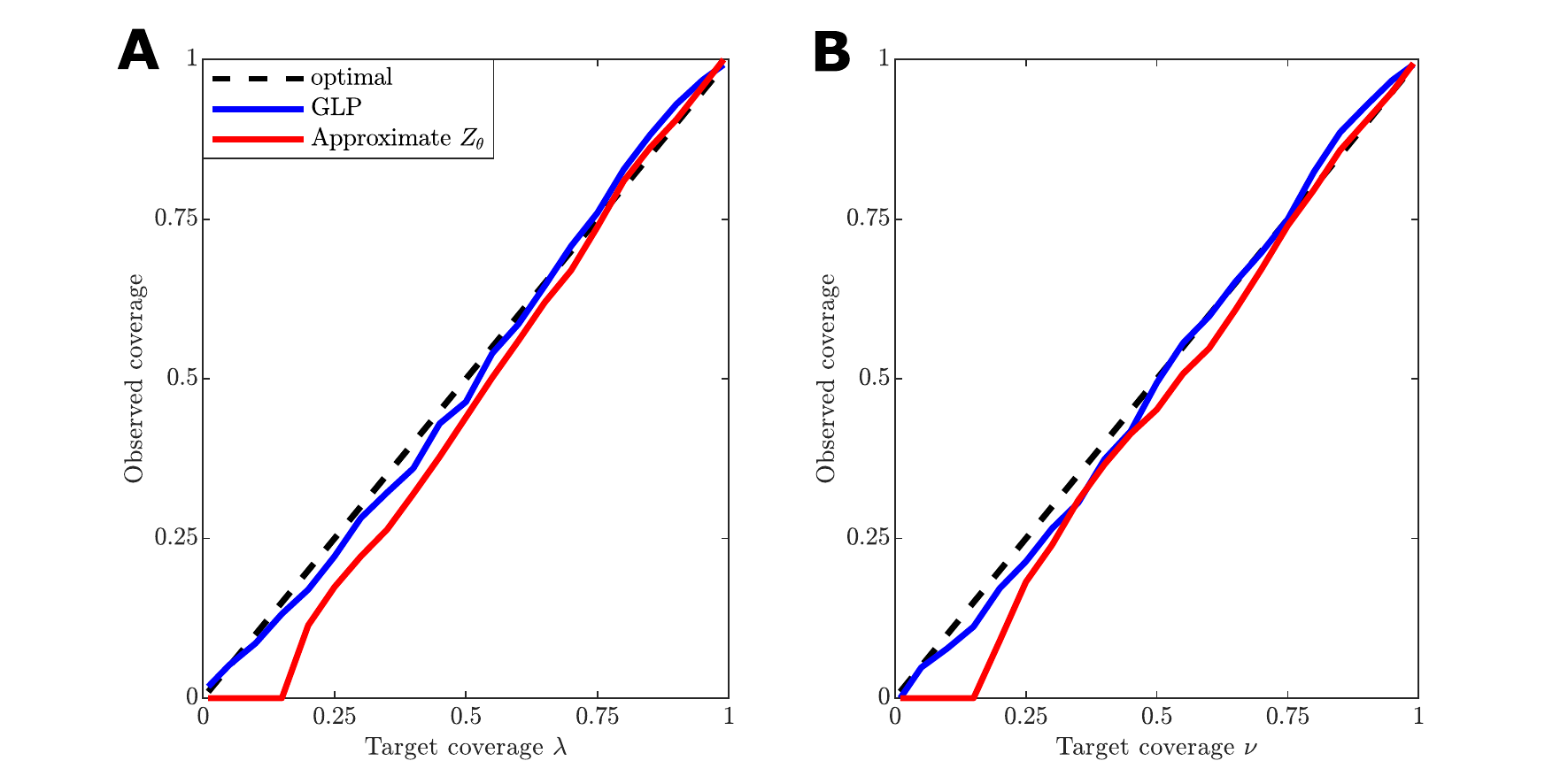}
	\caption{Coverage comparison between our calibration process (solid blue) and standard likelihood profile (solid red) using an accurate numerical approximation for $Z_{\boldsymbol{\theta}}$. Observed coverage is estimated using $B = 500$ independent datasets generated from the Conway--Maxwell-Poisson model using the true parameters.}
	\label{fig:coverageplotcmp}
\end{figure}

The resulting normalised profiles are shown in Figure~\ref{fig:prcmp}. Note that the width of the confidence intervals obtained under our GLP calibration method are wider than those under the true profile. This is reasonable, since the MGLE will not be as efficient as the MLE. However, even though the confidence intervals obtained with the generalised likelihood profiles are wider than those of the accurate approximation of the true MLE, they are obtaining the correct coverage as shown in Figure~\ref{fig:coverageplotcmp}. This accurate coverage property is a direct result of our calibration process~(Algorithm \ref{alg:calibdelta}).

%

\FloatBarrier

It is important to note that models like the CMP model are not the target application for our new method. Rather the types of problems that are the focus of this paper are when the likelihood is not available at all, not even point-wise up to a normalising constant. So there is no expectation that our method would outperform the standard MLE here, actually we would expect the opposite. The fact that confidence intervals are still reasonable for a tractable problem and that the observed coverage is very accurate demonstrates the utility of this approach for intractable models.

\FloatBarrier

\subsection{Stochastic model of biased diffusive transport}


We now consider a biased nearest-neighbour random walk with movement and death events on a one-dimensional interval $0 < x < L$ \citep{Ellery2012}. Space is discretised to give $N$ lattice sites with uniform spacing $\Delta x = L/(N-1)$, so that lattice site $i$ is located at position $x_{i} = (i-1)\Delta x$ for $i=1,\hdots,N$. Time is also discretised uniformly, into intervals of duration $\Delta t$, with time step $j$ occurring between times $t = (j-1)\Delta t$ and $t = j\Delta t$. An absorbing boundary applies at $x = 0$ (lattice site $1$) while a reflecting boundary applies at $x = L$ (lattice site $N$). Throughout, without loss of generality, we assume $\Delta t = 1$ and $\Delta x =1$ (i.e., $L = N-1$).

The random walk begins by releasing a particle at one of the $N$ lattice sites and ends when the particle leaves the simulation, either through absorption or a death event. During each time step, for internal lattice sites, $i = 2,\ldots, N-1$, the particle undergoes a movement event with probability $p_m$ or a death event with probability $p_d$ otherwise the particle remains at rest with probability $1-p_m - p_d$. If a movement event occurs the particle moves to the right with probability $p_r$ and to the left with probability $1-p_r$. The boundary lattice sites are dealt with differently (i) at the reflective boundary ($i = N$), a movement event is aborted with probability $p_r$, (ii) at the absorbing boundary ($i=1$), a movement event results in a particle is absorbed with probability one and removed from the simulation.

Let $T_{i}$ be the random variable denoting the lifetime of a particle released at position $x_i$. The $k$-th raw moment of $T_{i}$ is given by 
\begin{equation*}
\CondE{T_{i}^{k}}{\paramvec} = \sum_{j=0}^{\infty} t_{j}^{k} \CondPDF{t_{j}}{x_i,\paramvec},
\end{equation*}
where $\CondPDF{t_{j}}{x_i,\paramvec}$ is the probability mass that a particle released at lattice site $i$ is removed after $j$ time steps (i.e.~at time $t = j\Delta t$), and $\paramvec = (p_m,p_d,p_r)^\text{T}$ are the model parameters. Standard arguments \citep{Ellery2012} show that for small values of $\Delta x$ and $\Delta t$, $\CondE{T_{i}^{k}}{\paramvec}$ is well approximated by $M_{k}(x_{i},\paramvec)$, where $M_{k}(x,\paramvec)$ is a continuous function satisfying the boundary value problem:
\begin{equation}
D\ddydx{M_{k}(x,\paramvec)}{x} - v \dydx{M_{k}(x,\paramvec)}{x} - dM_{k}(x,\paramvec) = -k M_{k-1}(x,\paramvec), 
\label{eq:ivp}
\end{equation}
\begin{equation*}
	M_{k}(0,\paramvec) = 0,\quad \dydx{M_{k}(L,\paramvec)}{x} = 0,
\end{equation*}
with $M_{0}(x,\paramvec) = 1$ and the diffusivity, drift and decay coefficients identified as $D = (\Delta x)^{2}p_{m}/(2\Delta t)$, $v = \Delta x p_{m}(1- 2p_{r})/\Delta t$ and $d = p_{d}/\Delta t$, respectively.

Data from such a process will consist of $m$ particle lifetimes for $n$ unique initial particle positions, 
\begin{equation*}
\bvec{y} = \left[\begin{matrix}
t_{1}|x_1 & t_{2}|x_1 & \cdots & t_{m}|x_1 \\
t_{1}|x_2 & t_{2}|x_2 & \cdots & t_{m}|x_2 \\
\vdots & \vdots & \ddots & \vdots\\
t_{1}|x_n & t_{2}|x_n & \cdots & t_{m}|x_n 
\end{matrix}\right],
\end{equation*} 
where $t_i|x \sim \CondPDF{\cdot}{x}$ is the $i$th sample of the particle lifetime given initial position $x$. 
For our generalised likelihood construction, we apply the $r$-th order moment distance for the loss function,
\begin{equation}
\loss{\bvec{y}}{\paramvec} = \sum_{i=1}^n \sum_{k=1}^r \frac{(M_k(x_i,\paramvec) - \hat{M}_k(x_i,\bvec{y}))^2}{\hat{V}_k(x_i,\bvec{y})},
\label{eq:lossmoment}
\end{equation}
where
\begin{equation*}
\hat{M}_k(x_i,\bvec{y}) = \frac{1}{m}\sum_{j=1}^m t_{j}^k|x_i,
\end{equation*}
and $\hat{V}_k(x_i,\bvec{y})$ is a non-parametric bootstrap estimate of $\V{\hat{M}_k(x_i,\bvec{y})}$. Note that the moment distance in \eqref{eq:lossmoment} also falls within the minimum discrepancy framework of~\cite{Oates2022}.

Unlike the CMP model, stochastic models such as this are truly intractable. Importantly, the analytical expression for the boundary value problem~(\eqref{eq:ivp}) is only strictly valid for $p_r \approx 0.5$. Despite this, we show that our approach is robust and correctly highlights practical identifiablity problems for strong bias to the right.  

\subsubsection{Identifiable case: unbiased random walk $p_r = 0.5$}
We generate synthetic data of $n = 1,000$ particle lifetimes $\bvec{y} = [t_1, t_2, \ldots, t_n]$ through stochastic simulation with $p_m = 1$, $p_d = 0.001$ and $p_r = 0.5$. Then we treat $p_m$ as fixed and known, then perform our generalised likelihood profile approach for estimation of $p_d$ and $p_r$. We apply our MGLE and generalised likelihood profile approach to estimate 95\% confidence intervals for the death probability, $p_d$, and right-movement bias probability, $p_r$. 

\begin{figure}
	\centering
	\includegraphics[width=0.85\linewidth]{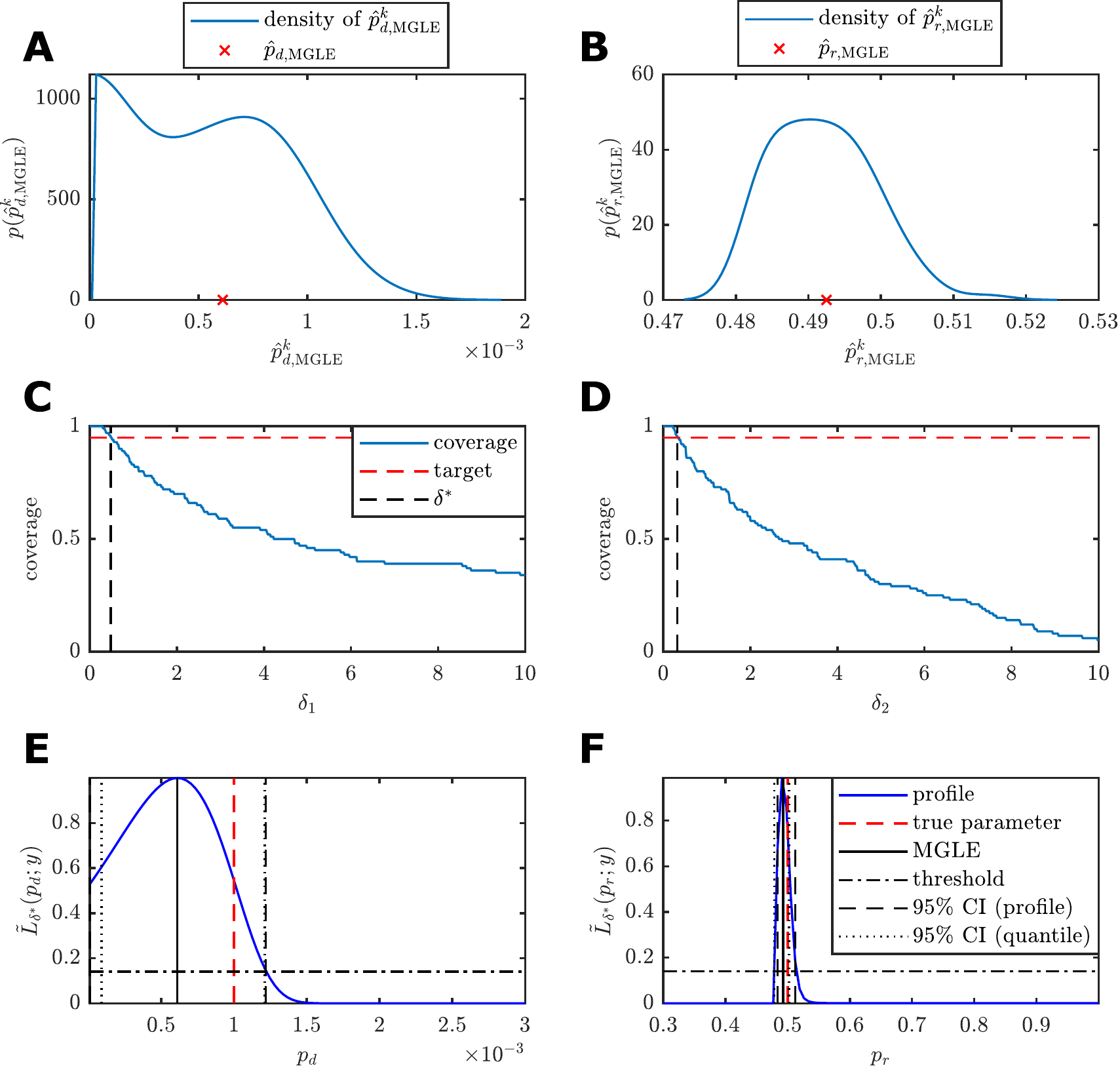}
	\caption{Results of generalised likelihood profiles and comparison with quantile-based bootstrap confidence intervals for the unbiased random walk case $p_r = 0.5$.  Distributions of the unscaled MGLE from bootstrap datasets for (A) the death parameter and (B) the right bias parameter. (C) and (D) Calibration plot for the scale parameters targeting  a coverage of 0.95. (E) and (F) Generalised likelihood profiles for each parameter with comparison between calibrated profile confidence intervals and quantile-based bootstrap confidence intervals.}
	\label{fig:prident}
\end{figure}

The results are shown in Figure~\ref{fig:prident}. In Figure~\ref{fig:prident}(A)--(B) the $K=100$ bootstrap MGLEs are shown to be concentrated around the MGLE related to the observed data indicating no bias from the moment distance loss function~(\eqref{eq:lossmoment}). The process of calibrating the generalised likelihood profiles is shown in Figure~\ref{fig:prident}(C)--(D) with optimal choices of $\delta^*$ indicated for both parameters of interest. Finally the calibrated profiles along with the 95\% CI are shown in Figure~\ref{fig:prident}(E)--(F). Quantile-based bootstrap confidence intervals are also indicated for comparison with a standard technique that is not based on the Wald interval. Such quantile-bootstrap confidence intervals are estimated by taking the $\alpha/2$ and $1-\alpha/2$ empirical quantiles of the MGLE samples, $\{\hat{\paramvec}^k_{\text{MGLE}}\}_{k=1}^{k=K}$. 

For both parameters, confidence intervals estimated by the quantile-based bootstrap  and our generalised lieklihood profile contain the true parameters. However, we note the quantile-based interval for $p_d$ does not contain zero, whereas our approach does. Given the distribution of the bootstrap MGLE samples, the containing of zero in the confidence interval seems appropriate. 
\FloatBarrier

\subsubsection{Non-identifiable case: biased random walk $p_r = 0.7$}
We now consider the case where the parameters of the data generating process lead to a practically unidentifiable problem for the right movement bias parameter $p_r$. We generate synthetic data of $n = 1,000$ particle lifetimes $\bvec{y} = [t_1, t_2, \ldots, t_n]$ through stochastic simulation with $p_m = 1$, $p_d = 0.001$ and $p_r = 0.7$ that represents a strong movement bias away from the absorbing left boundary. As with the unbiased case, we treat $p_m$ as a fixed and known parameter, then perform our generalised likelihood profile approach for estimation of $p_d$ and $p_r$. For such a strong bias to the right, very few particles reach the absorbing boundary. This leads to observations in which almost all particle removals are related to death events since few particles ever reach the absorbing boundary. This means we expect to obtain a precise estimate for the death probability $p_d$ but parameter non-identifiability in $p_r$.

\begin{figure}
	\centering
	\includegraphics[width=0.85\linewidth]{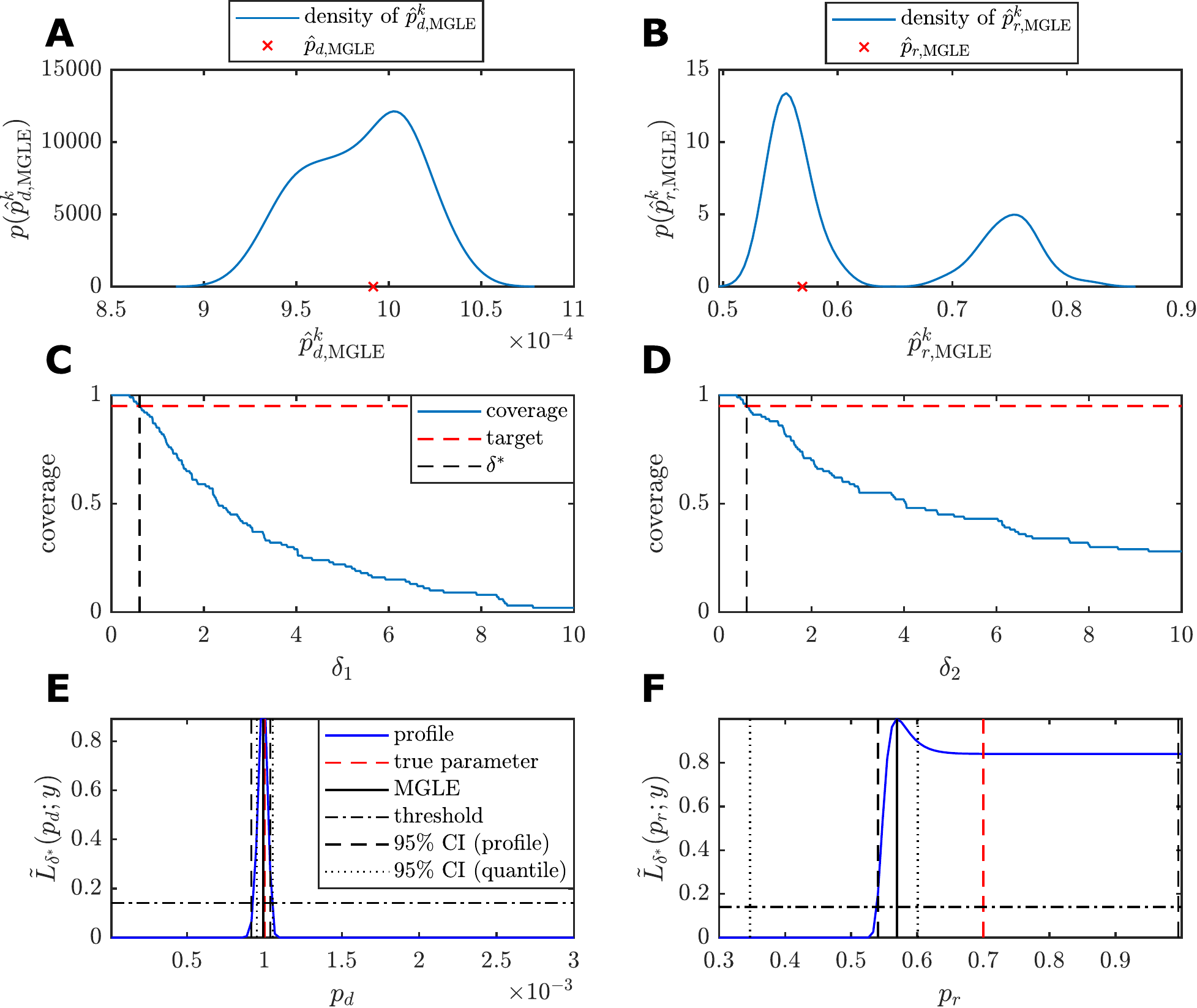}
	\caption{Results of generalised likelihood profiles and comparison with quantile-based bootstrap confidence intervals for the biased random walk case $p_r = 0.7$.  Distributions of the unscaled MGLE from bootstrap datasets for (A) the death parameter and (B) the right bias parameter. (C) and (D) Calibration plot for the scale parameters targeting  a coverage of 0.95. (E) and (F) Generalised likelihood profiles for each parameter with comparison between calibrated profile confidence intervals and quantile-based bootstrap confidence intervals.}
	\label{fig:prunident}
\end{figure}

The results are shown in Figure~\ref{fig:prunident}. In Figure~\ref{fig:prunident}(A)--(B) the $K=100$ bootstrap MGLEs are shown. We note that these are concentrated around the MGLE related to the observed data, however, this MGLE for $p_r$ does not match the true parameter at $p_r = 0.7$. However, there are a few bootstap MGLEs concentrated around the true parameter. This bi-modality is a result of the expected non-identifiability in $p_r$. The process of calibrating the generalised likelihood profiles is shown in Figure~\ref{fig:prunident}(C)--(D) with optimal choices of $\delta^*$ indicated for both parameters of interest. Finally the calibrated profiles along with the 95\% CI are shown in Figure~\ref{fig:prunident}(E)--(F). Quantile based bootstrap confidence intervals are also indicated for comparison. 

Unlike the unbiased case, confidence intervals estimated by the quantile-based bootstrap only contain $p_d$. Whereas our generalised likelihood profile contains both true parameters. Furthermore the practical non-identifiability of $p_r$ is strongly highlighted using our generalised profile approach with the profile flattening out on the right portion of the domain well above the 95\% CI threshold. Therefore, we conclude our method can be employed for the purposes of identifiability analysis for intractable likelihoods in the same way standard likelihood profiles are used when the likelihood is available~\citep{Raue2009}.

\FloatBarrier

\subsubsection{Observed coverage}

\begin{figure}[h]
	\centering
	\includegraphics[width=\linewidth]{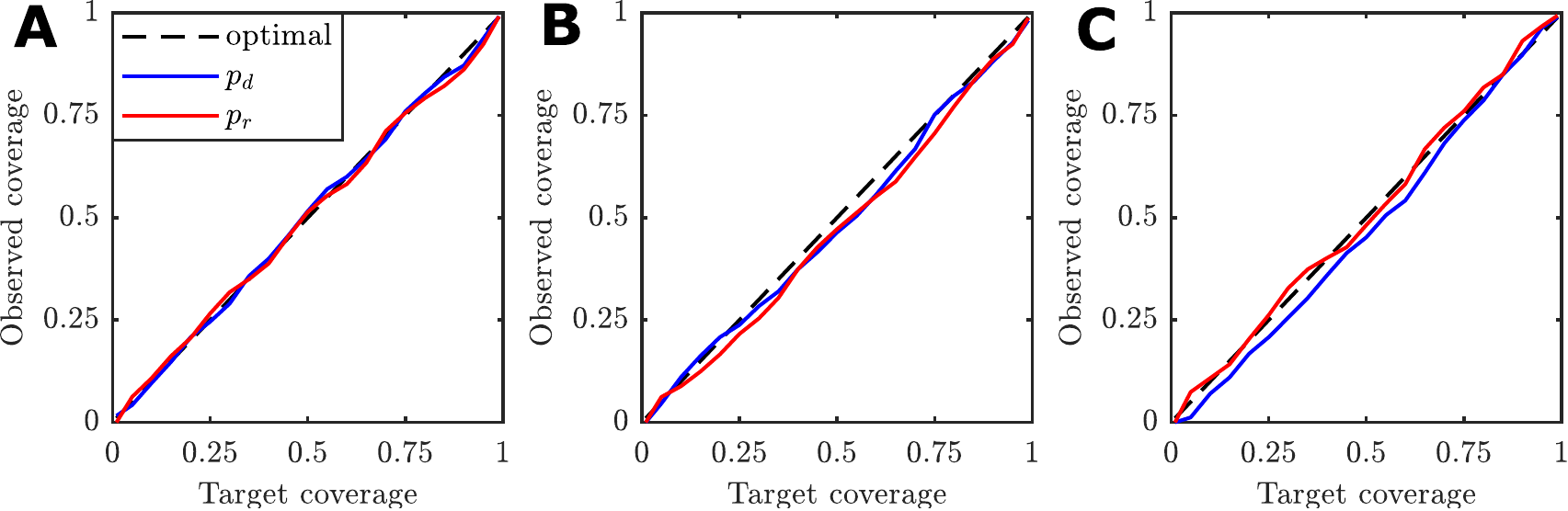}
	\caption{Coverage plots for our calibration process based on $B = 500$ independent datasets generated from the biased random walk model using the true parameters. True parameters are $p_m = 1$, $p_d = 0.001$, and (A) left movement bias $p_r = 0.48$, (B) unbiased movement $p_r = 0.5$, (C) right movement bias $p_r = 0.52$.}
	\label{fig:coverageplot}
\end{figure}

We also produce estimates of the observed coverage for our approach using data sets that are independent of the calibration procedure. This is done for both parameters $p_d$ and $p_r$ under several scenario for the true bias with $p_r \in [0.48, 0.5, 0.52]$. In each case, we achieve observed coverage that is extremely consistent with the target level of $1-\alpha$. These results, shown in Figure~\ref{fig:coverageplot}, along with those from the CMP model, demonstrates our approach obtains good frequentist properties for a range of true parameters and model settings. 

\FloatBarrier
\section{Discussion}
\label{sec:discuss}

In this work, we have developed a method for construction of profiles with similar properties of standard profile likelihood approaches, however, without the need of evaluating a likelihood function. Instead our approach is based on the concept of a generalised likelihood that requires an appropriate loss function be defined. We define this generalised likelihood with a single tuning parameter, $\delta$, that enables the profiles built upon the generalised likelihood to be calibrated toward a target frequentist coverage. This tuning process avoids the implicit reliance on an asymptotic assumption that is usually necessary. Through simulation studies, we show our approach is well behaved and obtains accurate coverage for repeated sampling over different models, true parameter sets, and datasets. Thereby we enable the use of likelihood profiles for parameter estimation, uncertainty quantification, and identifiablity analysis for stochastic models with intracable likelihoods that are ubiquitous in many fields of science. 

Beyond generalised likelihood profiling, our calibration approach also has potential value for generalised Bayesian inference~\citep{Bissiri2016,Matsubara2022}. For example, once the generalised likelihood has been calibrated, $\genlike{\bvec{y}}{\paramvec}{\delta^*}$, this, in turn could be applied as part of a Bayesian posterior sampling. This would avoid the computationally intensive methods currently applied to obtain Bayesian credible intervals in a generalised Bayesian setting~\citep{syring2019calibrating}, that involves obtaining many posterior approximations. Our approach could be uses as a pre-sampling conditioning of the generalised likelihood for the equivalent generalised Bayesian inference problem.

To validate our method, we test the coverage properties of our estimator for repeated bootstrap sample datasets generated with the true parameters $\paramvec_{\text{true}}$. However, for very complex models this may not be sufficient, and a modified calibration procedure could be required. A natural extension to ensure more robust calibration would be to specify a larger set of possible $\paramvec_{\text{true}}$ and calibrate $\delta^*$ in a minimax sense. That is, ensure the worst-case coverage is as close to the target as possible. In addition we could consider quantile regression approaches as used in the ACORE method~\citep{Dalmasso2020}. 

We also restricted our examples to use loss functions that lead to minimum discrepancy kernel estimators~\citep{matsubara2022generalised,Oates2022}. We show that the invariance of our estimator does not rely on this choice and applies for any loss function with a well defined minimum. It remains to be explored what asymptotic consistency or central limit theorem results could be obtained for more general loss functions. We suspect there could also be useful links made to the work surrounding the use of surrogate models with correction transformations~\citep{Bon2022,Warne2021}. This is a interesting line of future research that could further inform the calibration process for practical applications.  

Finally, we highlight the efficiency of our method. We only require a relatively small number of bootstrap samples to implement our calibration process, for example, $K = 100$ in all of our examples. Our approach is potentially more expensive than equivalent likelihood profiling when the likelihood is available, however, it is still substantially cheaper than the equivalent simulation-based inference approaches in the Machine Learning or Bayesian statistics literature. While these two statistical problems are not directly comparable, our calibration procedure offers a new approach for frequenstist statistics without likelihoods and provides insight into potential performance gains for Bayesian inference with intractable likelihoods. Ultimately our new statistical tools open the applicability of profile-based analysis to realistic intractable stochastic models and provides a practical approach to scientific inquiry.  

\paragraph{Acknowledgments}
This project was supported by the Australian Research Council (FT210100260, and DP230100025). DJW thanks Queensland University of Technology for support through the Early Career Researcher support scheme.

\paragraph{Data Availability}
Implementations of the generalised likelihood profile method and calibration are provide on GitHub at \href{https://github.com/davidwarne/GeneralisedLikelihoodProfiles}{https://github.com/davidwarne/GeneralisedLikelihoodProfiles} along with example usage scripts for the describe models in this manuscript.

\paragraph{Author contributions}
CD, MJS, and DJW designed the research. DJW, OJM, and EJC provided analytical tools. DJW, CD, MJS, and EJC provided computational tools. DJW performed the analysis. DJW, OJM, and CD analysed the results. All authors contributed to writing, revising and approving the final manuscript.

\bibliographystyle{apalike}


\end{document}